\documentclass{aastex63}

\renewcommand{\baselinestretch}{1} 

\usepackage{amsmath}
\usepackage{bm}
\usepackage{comment}

\graphicspath{{./}{figures/}}

\received{January 28, 2021}
\revised{August 11, 2021}
\accepted{August 16, 2021}
\submitjournal{ApJ}

\shorttitle{Merging Criteria for Planetesimal Collisions}
\shortauthors{Shibata et al.}

\begin{document}

\title{Merging Criteria for Planetesimal Collisions}

\correspondingauthor{Takashi Shibata}
\email{shibata.takashi@nao.ac.jp}

\author[0000-0002-0786-7307]{Takashi Shibata}
\affil{National Astronomical Observatory of Japan \\
2-21-1 Osawa, Mitaka, Tokyo 181-8588, JAPAN}
\affil{The University of Tokyo \\
7-3-1 Hongo, Bunkyo-ku, Tokyo 113-8654, JAPAN}

\author{Eiichiro Kokubo}
\affiliation{National Astronomical Observatory of Japan \\
2-21-1 Osawa, Mitaka, Tokyo 181-8588, JAPAN}
\affiliation{The University of Tokyo \\
7-3-1 Hongo, Bunkyo-ku, Tokyo 113-8654, JAPAN}

\author{Natsuki Hosono}
\affiliation{Center for Planetary Science \\
Integrated Research Center of Kobe University, 7-1-48 Minatojima-Minamimachi, Chuo-ku, Kobe 650-0047, JAPAN}
\affiliation{Japan Agency for Marine-Earth Science and Technology \\
3173-25 Showa-machi, Kanazawa-ku, Yokohama-city, Kanagawa 236-0001, JAPAN}
\affiliation{RIKEN \\
2-2-3 Minatojima-minamimachi, Chuo-ku, Kobe, Hyogo 650-0047, JAPAN}


\begin{abstract}
In the standard scenario of planet formation, terrestrial planets, ice giants, and cores of gas giants are formed by the accumulation of planetesimals.
However, there are few $N$-body simulation studies of planetesimal accretion that correctly take into account the merging condition of planetesimals.
In order to investigate a realistic accretion process of planetesimals, it is necessary to clarify the merging criteria of planetesimals at collision.
We perform numerical collision experiments using smoothed particle hydrodynamics and obtain the merging criteria as a function of planetesimal mass and impact parameters for undifferentiated rocky and icy planetesimals and differentiated icy planetesimals.
We vary the total mass of colliding planetesimals, their mass ratios, and the impact angle and obtain the critical impact velocity as the merging criteria distinguishing merging from hit-and-run collision.
We find that the critical impact velocity normalized by the two-body surface escape velocity decreases with increasing impact angle. 
The critical impact velocity does not depend on the total mass, while it has a weak positive dependence on the mass ratio.
These results barely depend on the composition and internal structure of the planetesimals.

\end{abstract}

\section{Introduction} \label{sec:intro}

Planets are formed from protoplanetary disks composed of dust and gas that exist around protostars \citep[e.g.,][]{1972epcf.book.....S, 1985prpl.conf.1100H}.
In the standard scenario, planetesimals first form from dust \cite[e.g.,][]{1973ApJ...183.1051G, 2002ApJ...580..494Y}. 
Terrestrial planets, giant planet cores, and ice giants are then formed mainly by the accumulation of planetesimals \cite[e.g.,][]{1985Sci...228..877W, 1998Icar..131..171K, 1998Icar..136..304C,1999Icar..142..219A, 2004ARA&A..42..549G}.
Recently, as an alternative model, pebble accretion model is proposed where the accretion of pebbles supplied by radial migration due to
gas drag accelerates the growth of planets \citep[e.g.,][]{2014A&A...572A.107L}.

The accumulation process of planetesimals has been studied by a statistical method based on the coagulation equation with kinetic theory \cite[e.g.,][]{1993Icar..106..190W, 2010Icar..209..836K} and $N$-body simulations \cite[e.g.,][]{1996Icar..123..180K}. 
The statistical method requires approximations to simulate the dynamical evolution of planetesimals. 
In particular, it uses a particle-in-a-box approximation for planetesimals, in other words, a uniform spatial distribution of planetesimals.
The statistical method can handle a huge number of planetesimals, which allows collisional fragmentation to be considered.
However, as planetesimal accretion proceeds, the spatial uniformity breaks down due to the emergence of runaway growing planetesimals \citep[e.g.,][]{1978Icar...35....1G, 1996Icar..123..180K}. 
On the other hand, $N$-body simulations can calculate the evolutions of mass, velocity, and spatial distributions simultaneously from first principles. 
However, since the calculation cost of the mutual gravity of planetesimals is very expensive, $N$-body simulations cannot be used for a large number of small planetesimals as a starting condition.
One of the largest scale {\em N}-body simulations to date  used one million planetesimals \citep{2019MNRAS.489.2159W}.
It is difficult to include collisional fragmentation in $N$-body simulations since the fragmentation increases the number of planetesimals.
Thus, most studies with $N$-body simulations assume perfectly inelastic collisions between planetesimals and do not consider bouncing and fragmentation.


Several calculations of the planetesimal accretion process have taken into account rebound and fragmentation \citep[e.g., ][]{2013Icar..224...43C, 2015ApJ...806...23L, 2017Icar..281..459M}. 
\citet{1990MNRAS.245...30B} used their own rebound and fragmentation model, dividing collisional outcomes into four regimes, accretion, rebound, rebound with cratering, and shuttering, according to the impact velocity of planetesimals.
Their collision model is too simple to distinguish the regimes properly since they did not consider the deformation and compression pressure of colliding planetesimals.
\citet{2015ApJ...806...23L} used their collision model obtained in \citet{2012ApJ...745...79L}.
\citet{2012ApJ...745...79L} obtained the merging criteria of planetesimals by studying the specific collision energy at which a colliding target loses half of its mass. 
Using the {\em N}-body code, PKDGRAV, they treated the target and projectile as rubble piles.
They categorized collisional outcomes into hit-and-run, merging, partial accretion, erosion, and super catastrophic disruption and determined the boundaries between those outcomes.
\citet{2017Icar..281..459M} used the merging criteria of differentiated rocky protoplanets obtained in \citet{2012ApJ...744..137G} (hereafter GKI12) for planetesimal collisions.
However, the merging criteria of undifferentiated planetesimals can be different from those of differentiated protoplanets. 
In addition, studies on collisions of icy planetesimals are limited \cite[e.g.,][]{1999Icar..142....5B}, and none have obtained the merging criteria that can be used in $N$-body simulations. 

The merging criteria play an important role in the planetesimal accretion process. 
If perfect merging is assumed, even a grazing collision leads to merging. 
Thus, planetesimals grow more rapidly than in reality.
In addition, the rotation of solid planets depends on the merging criteria of collisions. 
The rotation of terrestrial planets was investigated by {\em N}-body simulations of protoplanets with the merging criteria obtained in GKI12 that consider accretion and hit-and-run outcomes \citep{2010ApJ...714L..21K}.
They found that the spin angular momentum of planets becomes smaller than that obtained with perfect merging. 

The collisions of planetary bodies have been studied using Smoothed Particle Hydrodynamics (SPH) \cite[e.g., ][]{1992ARA&A..30..543M}, and the equation of state (EoS) is important for the fluid approximation. 
Since realistic planetesimals are not fluid, material properties such as material strength are also important.
In \citet{1999Icar..142....5B}, the importance of the material strength and self-gravity of rocky and icy planetesimals for susceptibility to breakage by a collision was investigated for a wide range of planetesimal sizes. 
Similarly, \citet{2015Icar..262...58G,2017Icar..294..234G} studied the collisions of rocky planetesimals using high-resolution SPH.
They obtained the criteria for erosion and fragmentation.

The giant impact of protoplanets has also been well studied by SPH \cite[e.g., ][]{2004ApJ...613L.157A, 2009AREPS..37..413A, 2009ApJ...700L.118M, 2010ApJ...719L..45M, 2012ApJ...744..137G}.
In particular, GKI12 reports collision experiments of protoplanets with a rocky mantle and an iron core and investigated their merging criteria.
They defined the critical impact velocity as the impact velocity that separates merging and non-merging and found it to be independent of the total mass of colliding protoplanets.
On the other hand, the critical impact velocity depends on the mass ratio and the impact angle of protoplanets. 
They derived the fitting formula of the critical impact velocity as a function of mass ratio and impact angle, which can be easily used in {\em N}-body simulations.

It is difficult to use the results of most of the previous studies on planetesimal collisions in $N$-body simulations because the merging criteria are not systematically represented by a simple formula like that in GKI12.
In addition, it is not clear whether we can use the merging criteria of GKI12 as a condition for planetesimals because they obtained this condition for protoplanets that are differentiated into a rocky mantle and an iron core.
The structure and composition of protoplanets is different from undifferentiated km-sized planetesimals.
It is difficult to distinguish between hit-and-run and merging for undifferentiated rocky/icy planetesimals, which are important collisional outcomes \citep{2012ApJ...745...79L}. 
Thus, the systematic merging criteria of planetesimal like that of the fitting formula of GKI12 are needed.

In this work, we perform numerical collision experiments of rocky and icy planetesimals using the SPH method to obtain the merging criteria for undifferentiated rocky and icy planetesimals. 
We also perform collision experiments of differentiated icy planetesimals.
Section 2 explains the model and calculation method.
In Section 3 we present the results of numerical collision experiments.
Section 4 is devoted to a summary and discussion.

\section{Numerical Method}

We perform SPH simulations of planetesimal collisions.
We vary the planetesimal mass and collision parameters and obtain the merging criteria as a function of the parameters.

\subsection{Planetesimal Models}

We assume spherical planetesimals with no rotation.
The compositions of rocky and icy planetesimals are assumed to be pure ice (${\rm H_2O}$) and basalt, respectively. 
Typical densities of rocky and icy planetesimals are about $2.7{\rm g/cm^{3}}$ and $1{\rm g/cm^{3}}$, respectively.
We adopt the Tillotson EoS, which is widely used in previous studies \citep[e.g., ][]{1999Icar..142....5B, 2012ApJ...744..137G, 2019Icar..321.1013E}.
The material parameters shown in \citet{1989icgp.book.....M} and \citet{1999Icar..142....5B} are used in the Tillotson EoS (see Appendix for details of the EoS).
We set the specific internal energy of the initial planetesimal to be $u = 1.0\times 10^{4}$J/kg in the standard model.
To evaluate the effect of the initial internal energy, we also adopt $u = 1.0\times 10^{5}$J/kg.

We also consider differentiated icy planetesimals with a rocky core and an icy mantle.
The mass ratio between rock and ice in icy planetesimals is uncertain. 
Following the previous studies \citep[e.g.,][]{2003ApJ...591.1220L, 2018ApJ...855...60H}, we assume a ratio of unity.
This ratio is consistent with the observations of comets \citep[e.g.,][]{2002EM&P...89..179H, 2005Sci...310..281K} and the theoretical model of icy planetesimal formation \citep{2014A&A...570A..35M}.

We neglect the material strength and internal friction of
planetesimals and regard planetesimals as self-gravitating fluid for
simplicity. We deal with 100 km-sized planetesimals in
the gravitational regime where the celestial body is dominated by the
self-gravity \citep{1999Icar..142....5B}. The gravitational regime for icy
and rocky planetesimals is over 100 m in radius. The material
strength, internal friction and other factors such as porosity can be
important for collisional outcomes. The model limitations will be
discussed in Section 4.

\subsection{Collision Parameters}

We perform more than 1,000 sets of collision experiments with different initial conditions.
We consider a target (larger planetesimal) with mass $M_\mathrm{tar}$ and radius $R_\mathrm{tar}$ and a projectile (smaller planetesimal) with $M_\mathrm{pro}$ and $R_\mathrm{pro}$.
We systematically vary the mass ratio $ \gamma = M_{\rm pro} / M_{\rm tar} $ as $\gamma$ = 1, 2/3, 1/2, 1/4, 1/10, and 1/20. 
The smallest planetesimal mass is given by $M_{\rm p} = 7.0\times 10^{-7}M_{\oplus}, 1.9\times 10^{-6}M_{\oplus}$ and $ 1.0\times 10^{-6}M_{\oplus}$ for icy, rocky, and differentiated planetesimals with radius $R_{\rm p} = 100$ km, respectively.
To investigate the dependence on the total mass $M_{\rm tot} = M_{\rm pro} + M_{\rm tar}$, in the case of icy planetesimals with $\gamma$ = 1, 1/4, and 1/10, we consider two further total masses that are two and five times the standard mass.

The impact velocity is the relative velocity between a target and a projectile $v_{\rm imp} = |\bm{v}_{\rm pro}-\bm{v}_{\rm tar}|$.
We vary the impact angle in the range $\theta$ = 15--75 deg in 15 deg steps, and the impact velocity in the range $v_{\rm imp} =$ 1.0--3.0 $v_{\rm esc}$, where $v_{\rm esc}$ is the two-body surface escape velocity defined as $ v_{\rm esc}=\sqrt{2GM_{\rm tot}/(R_{\rm tar} + R_{\rm pro})}$.
The impact velocity is varied with a step size of about 0.4$v_{\rm esc}$ to find the boundary between merging and non-merging. 
Around the boundary we reduce the step size to 0.02$v_{\rm esc}$ to find it precisely.
The escape velocity ranges from 123 ms$^{-1}$ to 878 ms$^{-1}$ in our model.

\begin{figure}  
\includegraphics[width=\textwidth]{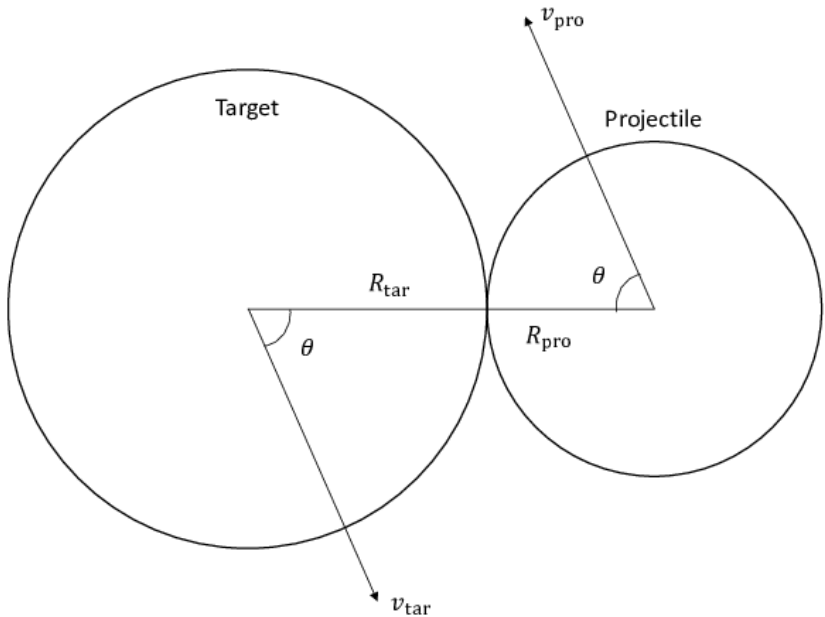}
\caption{
Geometry of a collision between target and projectile planetesimals with radii $R_{\rm tar}$ and $R_{\rm pro}$ and velocities $v_\mathrm{tar}$ and $v_\mathrm{pro}$ where $\theta$ is the impact angle.
\label{coll_pic}}
\end{figure}

The impact parameters $v_{\rm imp}$ and $\theta$ are defined when the two planetesimals are in contact with each other (Fig. \ref{coll_pic}).
To determine the impact parameters at the time of collision, by assuming a planetesimal to be a point mass, the trajectory of the projectile is calculated backwards as a hyperbolic trajectory, and the point $3(R_{\rm tar}+R_{\rm pro})$ away from the target is set as the initial position of the projectile.  

\subsection{Simulation Method \label{Particles}}

We use standard SPH \citep{1992ARA&A..30..543M}, parallelized via the framework for developing particle simulators (FDPS) \citep{2016ascl.soft04011I}.
We use the Wendland $C^6$ kernel introduced in \citet{doi:10.1111/j.1365-2966.2012.21439.x} for the kernel function, and adopt the artificial viscosity formula of \citet{1997JCoPh.136..298M} and \cite{2016ApJS..224...32H}.  
For the time integration, the leap-frog method is used.
For details of the SPH code used here, see \citet{2017PASJ...69...26H}.

The number of SPH particles of a projectile with $R_{\rm p} = 100$ km and $M_{\rm p}$
is 10,000 for all compositions and that of a target is changed in
proportion to its mass, which keeps the particle resolution constant
throughout the calculation.

We also simulate models with projectiles with 30,000 and 50,000 particles to investigate the dependence on the number of particles.
However, we find that their collisional outcomes barely change.
Therefore, considering calculation costs, we adopt 10,000 particles for $M_\mathrm{p}$ in this study. 

The initial planetesimals are created by the relaxation method.
First, SPH particles are arranged in a three-dimensional lattice pattern inside of a spherical region with the radius of a planetesimal. 
We set the initial specific internal energy for each SPH particle and start relaxation of particles.
SPH particles oscillate around the initial position without changing the center of mass of the spherical planetesimal. 
We wait for the velocity of SPH particles to become less than $5\%$ of the surface escape velocity of the planetesimal.

We also adopt the same initial conditions as GKI12 and compare the results to check the validity of our method.

\subsection{Collisional Outcomes}

In previous works \cite[e.g.,][]{2010ChEG...70..199A,2010ApJ...714.1789L,2012ApJ...745...79L}, the collisional outcomes of solid objects were classified into various categories, such as merging, hit-and-run, fragmentation, partial accretion, erosion, and cratering.
Of those categories, merging and hit-and-run are the most important outcomes in planetary accretion \cite[e.g.,][]{2010ApJ...714L..21K, 2012ApJ...745...79L} 
and therefore, in this study, we focus on merging and hit-and-run, following \cite{2012ApJ...744..137G}.

We simulate a collision for $10^5$ s and then analyze the outcome in the same way as \cite{2004ApJ...613L.157A} and \cite{2012ApJ...744..137G}.
In order to judge merging or non-merging, we use the mass of the largest remnant $M_{\rm lr}$ after the collision.
To identify bound objects, we use the friends-of-friends algorithm of \citet{1982ApJ...257..423H}.
We find the largest and the second largest objects. 
If their relative velocity exceeds their escape velocity, the larger object becomes the largest remnant.
If it falls below the escape velocity, the mass of the largest remnant is the total mass of the two objects.
The former is regarded as hit-and-run, while the latter is merging.

\section{Results}

\subsection{Merging and Hit-and-Run}

\begin{figure}  
\includegraphics[width=\textwidth]{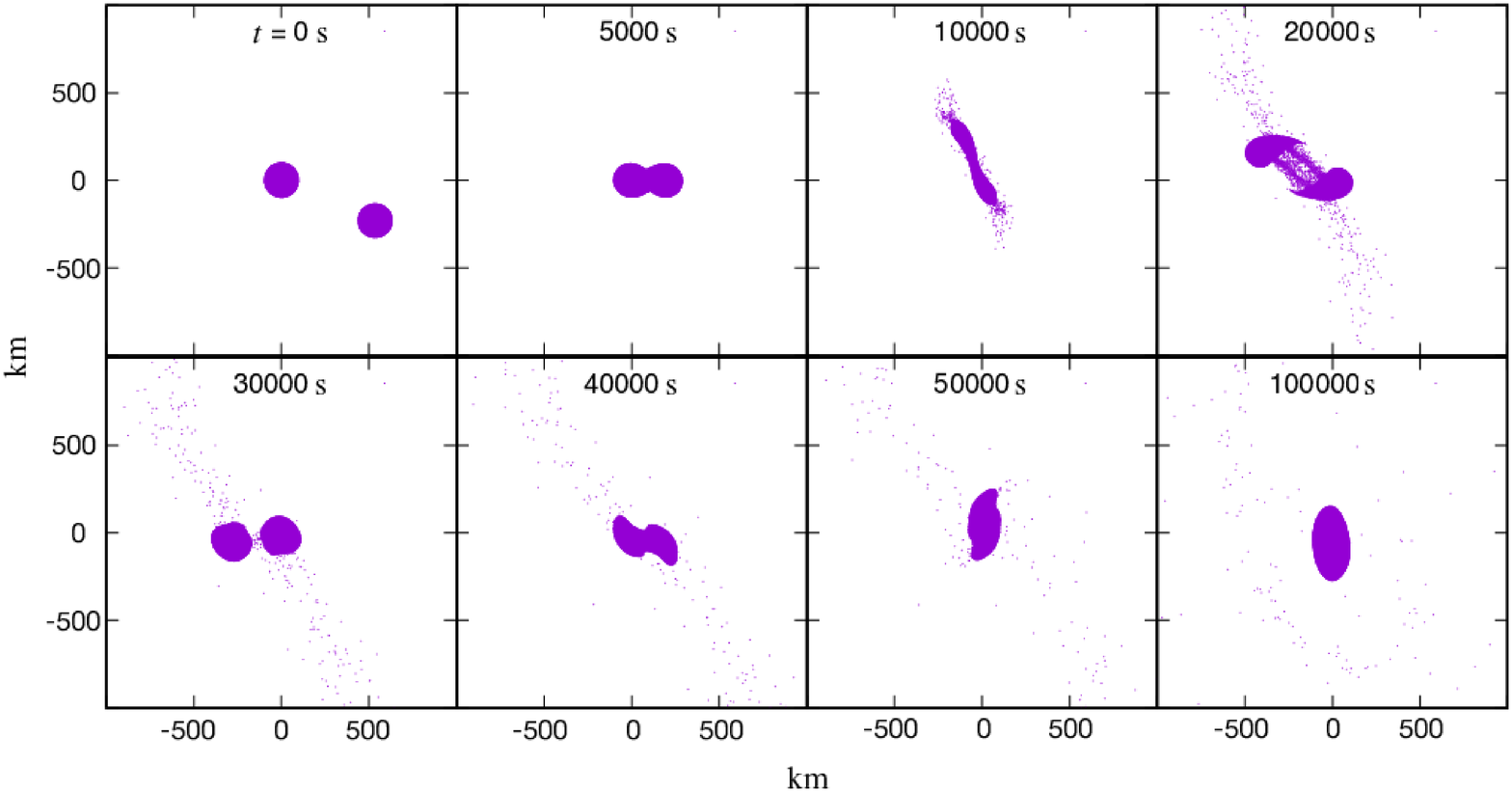}
\caption{
Snapshots of a typical merging collision between equal-mass icy planetesimals with mass $M_{\rm p} = 7.0\times 10^{-7}M_{\oplus}$ and radius $R_{\rm p} = 100$ km with 10,000 particles.
The impact parameters are $v_{\rm imp}$ = 1.3$v_{\rm esc}$ and $\theta = 30^\circ$.
\label{snapm}}
\end{figure}

\begin{figure}  
\includegraphics[width=\textwidth]{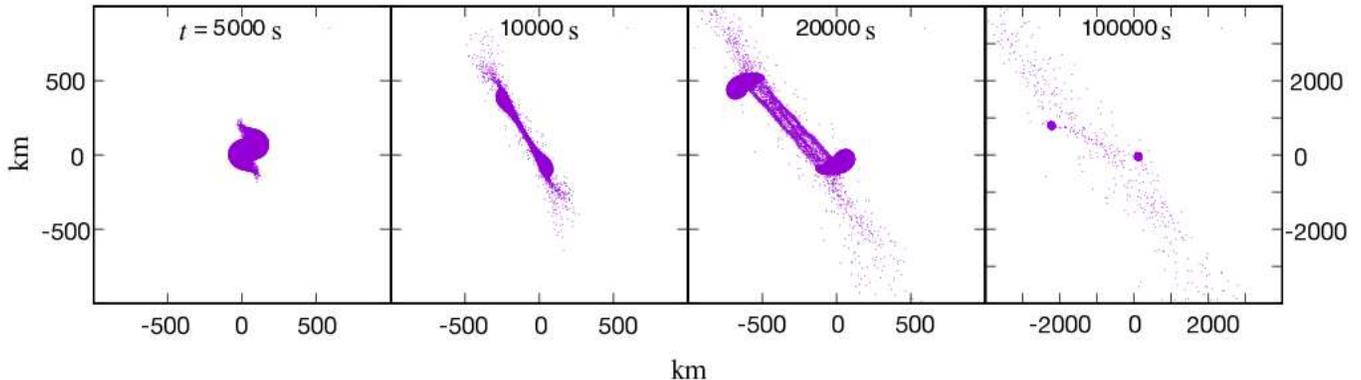}
\caption{
Snapshots of a typical hit-and-run collision between equal-mass icy planetesimals with mass $M_{\rm p} = 7.0\times 10^{-7}M_{\oplus}$ and radius $R_{\rm p} = 100$ km with 10,000 particles. 
The impact parameters are $v_{\rm imp}$ = 1.7$v_{\rm esc}$ and $\theta = 30^\circ$.
\label{snaph}}
\end{figure}

In the present paper, we only consider merging and hit-and-run collisional outcomes. 
Figures \ref{snapm} and \ref{snaph} show examples of the two outcomes for identical icy planetesimals with mass $M_{\rm p} = 7.0\times 10^{-7}M_{\oplus}$.
Figure \ref{snapm} shows the time evolution of a collision with a relatively small impact velocity
 $v_{\rm imp}$ = 1.3$v_{\rm esc}$ and an impact angle $30^\circ$.
After the first collision, the planetesimals are pulled back by mutual gravity and eventually merge.
Though some planetary surface materials are scattered around, most of the material is eventually reaccreted by the planetesimal.
With a slower impact velocity, planetesimals merge directly without rebound.
These types of collisions are considered as merging. 

On the other hand, a collision with a relatively higher impact velocity results in a hit-and-run.
In Figure \ref{snaph}, we show the result for a collision with the same condition as Figure \ref{snapm} but for a higher impact velocity $v_{\rm imp}$ = 1.7$v_{\rm esc}$.
After the collision, the planetesimals separate from each other with a velocity higher than the two-body escape velocity.
We regard this type of collisions as a hit-and-run.

Using the results at $t=10^{5}$ s, we judge the collisional outcome. 
In some calculations, at $t=10^{5}$ s there are two planetesimals geometrically separated with a relative velocity smaller than their two-body escape velocity.
We regard this case as merging since they are expected to merge eventually.

\subsection{Critical Impact Velocity \label{total_mass} \label{Energy}}

\subsubsection{Definition}

\begin{figure}  
\includegraphics[width=\textwidth]{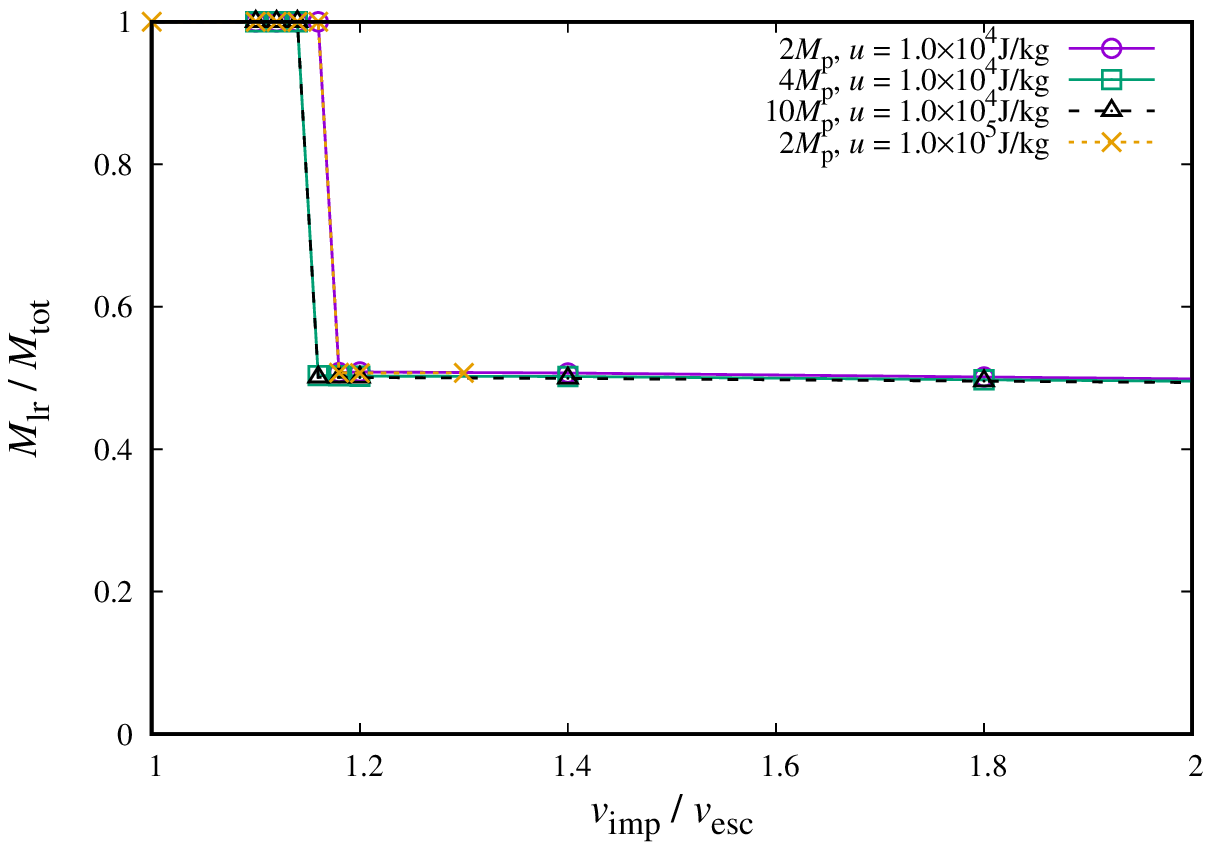}
\caption{
Mass of the largest gravitationally bound object normalized by the total mass, $M_{\rm lr} / M_{\rm tot}$, as a function of the impact velocity normalized by the two-body surface escape velocity, $v_{\rm imp}/v_{\rm esc}$ for impacts between two equal-mass icy planetesimals with $\theta = 60 ^\circ$ where the total mass and specific internal energy $u$ of planetesimals are 2, 4 and 10 $M_{\rm p}$ with $M_{\rm p} = 7.0\times 10^{-7}M_{\oplus}$ and $1.0 \times 10^4$ and $1.0 \times 10^5$ J/kg, respectively. 
\label{fig:largestremnant}}
\end{figure}

Figure \ref{fig:largestremnant} shows the mass of the largest body normalized by the total mass of colliding bodies, $M_{\rm lr}/M_{\rm tot}$, with respect to the impact velocity normalized by the two-body surface escape velocity, $v_{\rm imp}/v_{\rm esc}$,  for icy planetesimals  of different total masses with $\gamma = 1$ and $\theta =  60^\circ$.
There exists a discontinuous change in $M_{\rm lr} / M_{\rm tot}$ near $v_{\rm imp} = 1.16v_{\rm esc}$.
For an impact velocity less than this value, the two planetesimals merge; in other words, $M_{\rm lr} / M_{\rm tot} \simeq 1.0$.
For a larger impact velocity, the two planetesimals experience a hit-and-run and $M_{\rm lr} / M_{\rm tot} \simeq 0.5$, where most of the mass is separated into the target and the projectile and a part of the surfaces of the planetesimals is scraped off.
The critical value $v_{\rm imp} = 1.16v_{\rm esc}$ that separates merging from hit-and-run is referred to as the critical impact velocity $v_{\rm cr}$.

However, in the case of a nearly head-on collision ($\theta = 15^\circ$), there is no large change in $M_{\rm lr}/M_{\rm tot}$, but instead it changes smoothly with the impact velocity.
\citet{2012ApJ...745...79L} calls such collisions with $M_{\rm lr} \geq M_{\rm tar}$ partial accretions, while the other outcomes of these types of collisions are erosions. 
They showed that in general as the impact velocity increases, the collisional outcome changes from merging to partial accretion to erosion.
Note that our high-speed impact calculations are consistent with \citet{2012ApJ...745...79L}. 
Erosion mostly occurs for unrealistically high velocity collisions ($v_{\rm imp} > 3v_{\rm esc}$) which are not seen in planetesimal accretion processes \citep{2012ApJ...745...79L}.
Since the probability of nearly head-on high-velocity collisions
is quite low during the planetesimal accretion process, their
collision outcomes barely affect the accretionary evolution (GKI12).

\subsubsection{Total Mass Independence}

\citet{2010ChEG...70..199A} showed that when a planetesimal is in the gravitational regime, the critical impact velocity does not depend on the total mass of the colliding planetesimals. 
We confirm this total-mass independence of the critical impact velocity.
In Figure \ref{fig:largestremnant}, as shown in GKI12, $v_{\rm cr} / v_{\rm esc}$ barely depends on the total planetesimal mass. 
In fact, for $M_{\rm tot} = 2M_{\rm p}, 4M_{\rm p}$, and $ 10M_{\rm p}$, $v_{\rm cr} / v_{\rm esc} = 1.16$, $1.14$, and $ 1.14$, respectively.

For the impact velocities considered here, the second term is dominant for compression in the Tillotson EoS for the condensed region (Eq. \ref{condensed}) and no phase transition occurs by collisions.
In the case where planetesimals bounce back by the compressive pressure, which barely depends on the composition, the collisional outcome barely depends on the total planetesimal mass with a fixed target-to-projectile mass ratio \citep{2010ChEG...70..199A}. 
In other words, the critical impact velocity increases with the planetesimal mass and $v_{\rm esc}$ also increases likewise.
Then, $v_{\rm cr} / v_{\rm esc}$ is independent of the planetesimal mass.
GKI12 also showed that $v_{\rm cr} / v_{\rm esc}$ barely depends on the total mass for protoplanet collisions. 

Under the assumption of the self-gravitating fluid planetesimals the collision outcome has little dependence on the total mass.
However it should be noted that the material strength, internal
friction, porosity and other factors of planetesimals may affect this
dependence, particularly for small planetesimals.

\subsubsection{Internal Energy Independence}

To investigate the initial internal energy dependence of the critical impact velocity, several calculations are performed while varying the internal energy.
In a high-pressure situation, the compression pressure becomes more important than the thermal pressure. 
For this reason, it is expected that the pressure calculated for a collision is mostly independent of the initial internal energy.
We compare the critical impact velocity for $u = 1.0\times 10^{4}$ and $1.0\times 10^{5}$J/kg in  Fig. \ref{fig:largestremnant}.
We use equal-mass icy planetesimals and change the impact angle ($\theta = 15 $--$ 75^\circ$). 
For a given impact angle, no significant change is seen in the critical impact velocity.
For example, for $\theta = 15^\circ$ and $\theta = 60^\circ$, $v_{\rm cr} / v_{\rm esc} = 2.26$ and $1.16$, respectively, for both values of specific internal energy $u = 1.0\times 10^{4}$ and $1.0\times 10^{5}$J/kg (Figure \ref{fig:largestremnant}). 
Therefore, the dependence of the critical impact velocity on the internal energy of the initial planetesimal can be ignored, as shown in GKI12.

\subsection{Merging Criteria of Icy Planetesimals}

We investigate the basic dependencies of the critical impact velocity on the collision parameters.

\subsubsection{Dependence on Impact Angle \label{Impact_Angle}}

\begin{figure}  
\includegraphics[width=\textwidth]{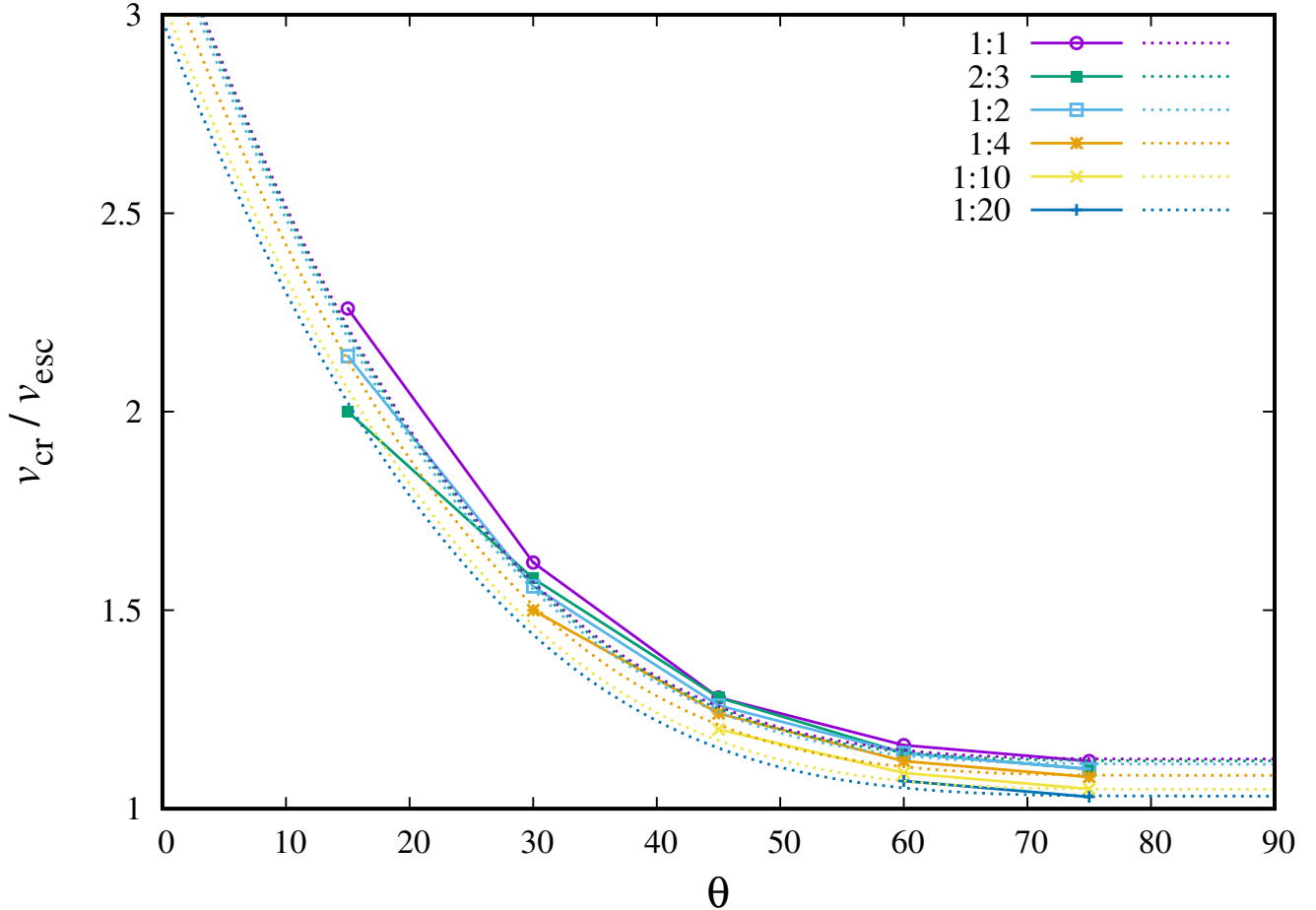}
\caption{
Normalized critical impact velocity $v_{\rm cr}/v_\mathrm{esc}$ against the impact angle $\theta$ for $\gamma = 1$, $2/3$, $1/2$, $1/4$, $1/10$, and $1/20$ for the standard planetesimal model.
The dotted curves represent the fitting formula Eq. \ref{fitting}.
The different colors correspond to the different mass ratios. 
\label{fit}}
\end{figure}

The critical impact velocity depends on the impact angle. 
Figure \ref{fit} shows the critical impact velocity against the impact angle for collisions of equal-mass icy planetesimals. 
As an example, in the standard model with $M_{\rm tot} = 2M_{\rm p}$, $\gamma = 1$, and $u = 1.0\times 10^{4}$J/kg, we obtain
$v_{\rm cr}/v_{\rm esc} = 1.16$ and $2.26$ for $\theta = 60^\circ$ and $15^\circ$, respectively. 
We find that the critical impact velocity decreases with increasing impact angle \citep{2010ChEG...70..199A, 2012ApJ...744..137G}.
As impacts approach the head-on impact ($\theta = 0^\circ$), $v_{\rm cr}$ increases sharply. 
For small impact angles, planetesimals efficiently lose relative kinetic energy by the inelastic collision, which leads to merging.
At collisions with large impact angles the energy dissipation is less effective than those with small impact angles. 
In this case, the colliding planetesimals tend to hit-and-run.
This can be qualitatively explained in terms of the size of the overlapping volume of colliding planetesimals (GKI12). 
Since this is geometrically smaller for higher impact angles, the fraction of kinetic energy converted to thermal energy of planetesimals and kinetic energy of fragments is small, which leads to a hit-and-run collision. 

\subsubsection{Dependence on Mass Ratio \label{Mass_Ratio}}

\begin{figure}  
\includegraphics[width=\textwidth]{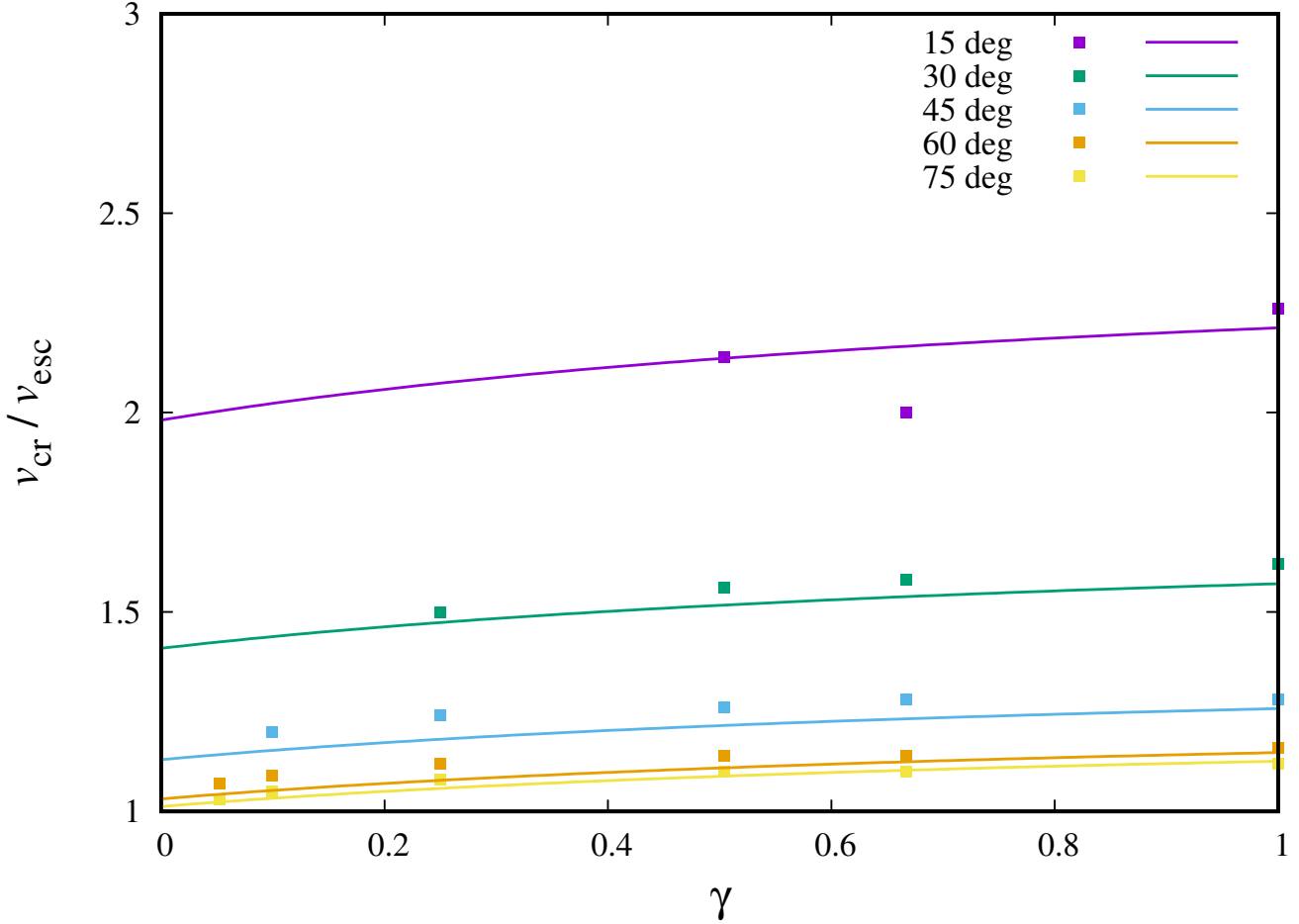}
\caption{
Normalized critical impact velocity $v_{\rm cr}/v_\mathrm{esc}$ plotted against the mass ratio $\gamma$. 
The different colors of the symbols and curves show the different impact angles. 
The solid curves represents the fitting formula of the critical impact velocity (Eq. \ref{fitting}).
\label{mass_ratio_dependece}}
\end{figure}

The critical impact velocity only weekly depends on the mass ratio $\gamma$.
The results for $\gamma = 1$, $2/3$, $1/2$, $1/4$, $1/10$ and $1/20$ with the standard planetesimal model are shown in Figure \ref{fit}.
We find that $v_{\rm cr} / v_{\rm esc}$ decreases slightly as $\gamma$ decreases. 
For example, for $\theta = 60^\circ$, $v_{\rm cr} / v_{\rm esc} = 1.16$ and $1.09$ for $\gamma = 1$ and $1/10$, respectively.
The change of the critical impact velocity is less than $10\%$ over $\gamma = 1/20$--1 (Fig. \ref{mass_ratio_dependece}).

\begin{figure}  
\includegraphics[width=\textwidth]{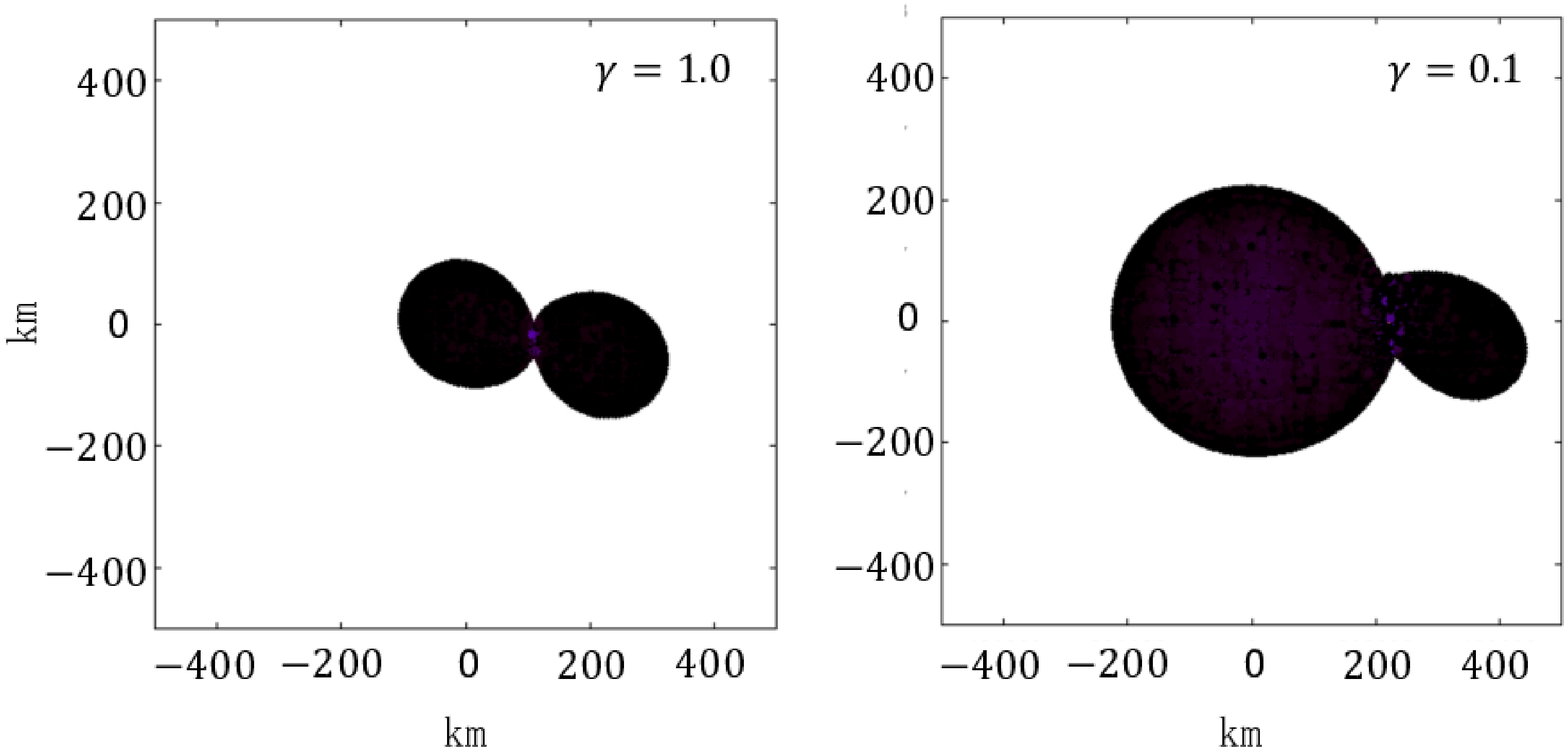}
\caption{
Snapshots of planetesimals at collision with $v_{\rm imp} = 1.1v_{\rm esc}$ and  $\theta = 60^\circ$for $\gamma = 1.0$ (left) and  $0.1$ (right).
\label{elip_snap}}
\end{figure}

Approaching planetesimals are deformed and become ellipsoids by the tidal force (Fig. \ref{elip_snap}).
When $\gamma$ is near unity, the target and projectile are deformed almost equally.
In a collision with $v_{\rm imp} = 1.1v_{\rm esc}$ and $\theta = 60^\circ$, the minor-to-major axis ratio of the bodies is 0.85 for $\gamma = 1$ when they collide.
The actual collision angle is $\simeq 52^\circ$ and the collision results in merging.
When $\gamma$ is smaller, although the projectile is deformed, the target is not deformed significantly. 
Under the same collision conditions but with $\gamma = 0.1$, the axis ratio is about 0.80 for the projectile and 1.0 for the target, respectively, and the actual collision angle is $\simeq 58^\circ$. 
The colliding planetesimals hit and run.
We find that the target deformation changes the actual collision angle.
The actual collision angle increases with decreasing $\gamma$, which reduces the collisional cross-section of the deformed bodies.
Thus, less energy dissipation takes place for smaller $\gamma$ and collisions with smaller $\gamma$ result in hit-and-run outcomes.

\subsubsection{Fitting Formula}

The critical impact velocities for the other impact angles that are not calculated in this study can be determined by an interpolation using the results above.
Following GKI12, using the least-square-fit method we derive the formula of the critical impact velocity as
\begin{equation}
    \frac{v_{\rm cr}}{v_{\rm esc}} = c_1\Gamma^2 \Theta^{c_5} + c_2\Gamma^2 + c_3\Theta^{c_5} + c_4, 
    \label{fitting}
\end{equation}
where $\Gamma = (M_{\rm tar} - M_{\rm pro})/M_{\rm tot} = (1-\gamma)/(1+\gamma)$ and $\Theta = 1 - $sin$\theta$. 
We obtain $c_1 = -0.232$, $c_2 = -0.114$, $c_3 = 2.14$, $c_4 = 1.12$, and $c_5 = 2.27$.
Note that the coefficients for all planetesimal models are summarized in Table \ref{table1}.
This fitting is shown by the dotted curves in Figure \ref{fit}.
Below this curve, colliding planetesimals merge, while above this curve they experience a hit-and-run.

The largest difference between the fitting formula and the plots is $\sim 0.2v_{\rm esc}$ at $15^\circ$ for $\gamma = 2/3$.
However, since head-on collisions are rare in planetesimal accretion, this difference is barely significant in the accretion process.
For $\theta \geq 45^\circ$, the fitting formula agrees well with the numerical results.

\subsection{Dependence on Composition and Differentiation}

\subsubsection{Composition}

\begin{figure}  
\includegraphics[width=\textwidth]{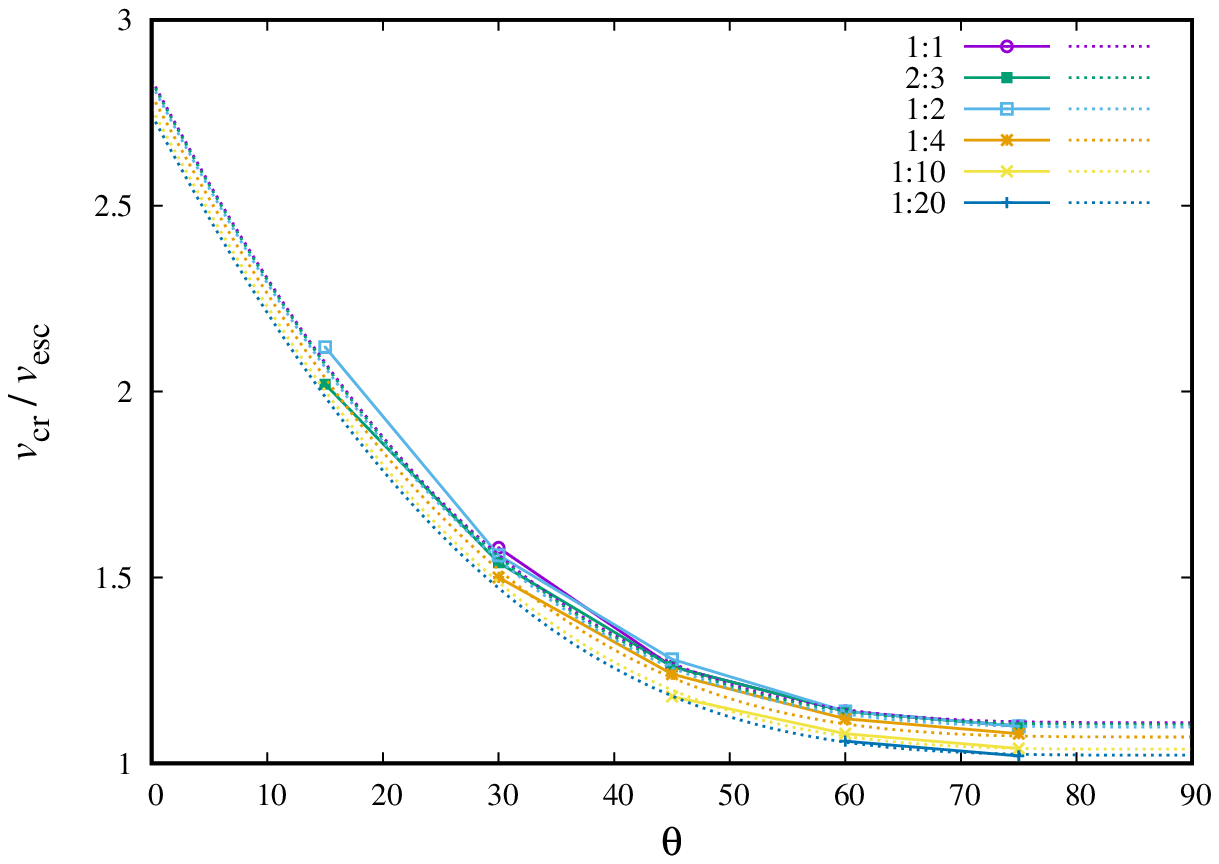}
\caption{As Figure \ref{fit} but for rocky planetesimals.
The dotted curves represent the fitting formula Eq. (\ref{fitting}). The different colors correspond to the different mass ratios.
\label{Rock}}
\end{figure}

We investigate the merging criteria of 100 km-sized rocky planetesimals in the same way as for icy planetesimals. 
As can be seen in Figure \ref{Rock}, there is no significant difference in $v_{\rm cr}$ due to composition difference. 
For the rocky planetesimal, the fitting coefficients are
$c_1 = -0.00863$, $c_2 = -0.107$, $c_3 = 1.73$, $c_4 = 1.11$, and  $c_5 = 1.94$.

The shape of colliding planetesimals is determined by their gravity when they approach.
The tidal deformation depends on their density ratio \citep[e.g.,][]{2010ChEG...70..199A}.
Since rocky and icy planetesimals are weakly compressible, their density ratios barely change during approaching and are almost unity independently of composition.
Therefore, the tidal deformation takes place in a similar manner for both planetesimals, which determines the critical impact velocity.
Since the density of icy and rocky planetesimals do not increase significantly at collision, the pressure due to collisional compression barely depends on the composition.
In addition, since no phase transition of planetesimals occurs by collision, the vaporized substance does not contribute to the bounceability.
For these reasons, it is not surprising that rocky and icy planetesimals have similar $v_{\rm cr}$.

\subsubsection{Differentiation}

\begin{figure}  
\includegraphics[width=\textwidth]{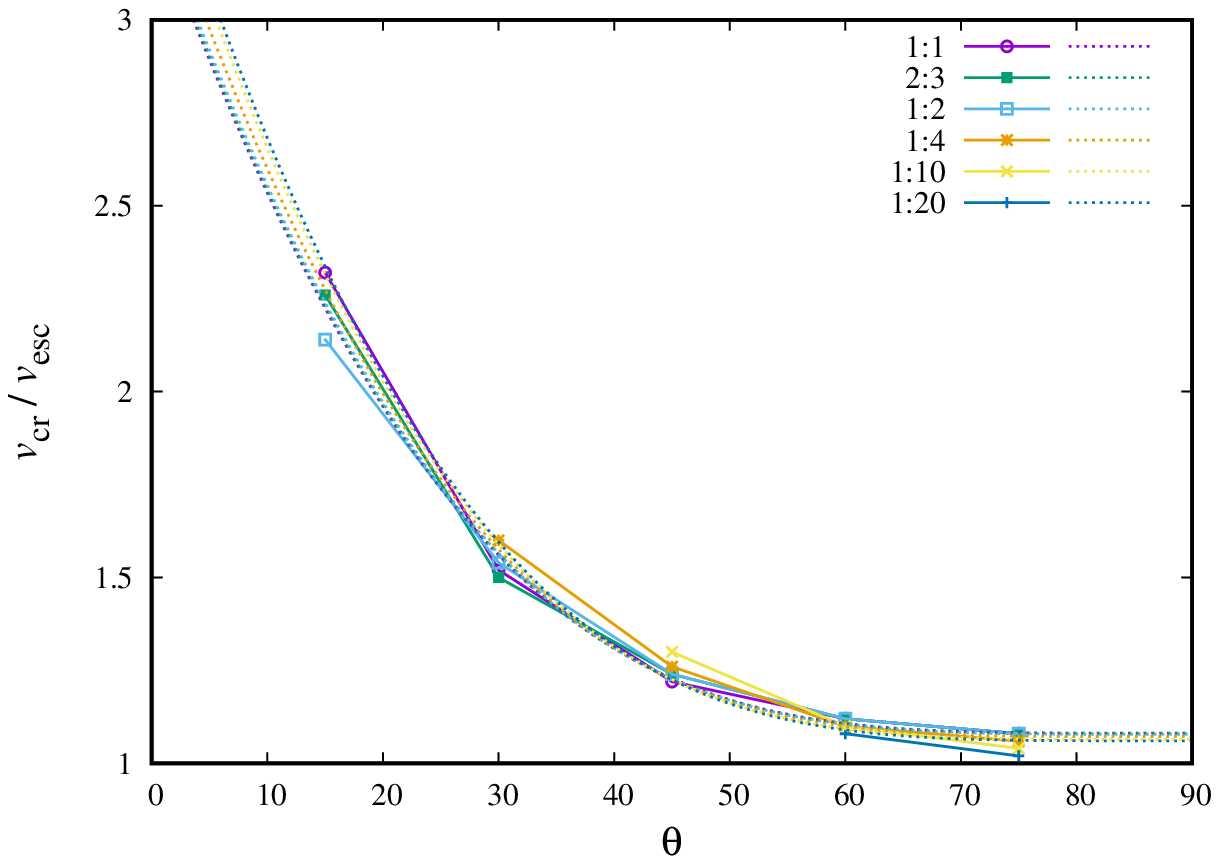}
\caption{As Figure \ref{fit} but for differentiated icy planetesimals.
The dotted curves represent the fitting formula Eq. (\ref{fitting}). The different colors correspond to the different mass ratios.
\label{differentiated}}
\end{figure}

Figure \ref{differentiated} shows the critical impact velocity as a function of impact angle for 100 km-sized differentiated icy planetesimals. 
We find that $v_{\rm cr}$ is not affected by the presence of the core.  
We obtain the fitting coefficients $c_1 = 0.313$, $c_2 = -0.0249$, $c_3 = 2.21$, $c_4 = 1.08$, and $c_5 = 2.21$.

Regardless of whether or not a core exists, all head-on collisions with reasonable velocities result in merging.
On the other hand, the target's mantle affects the projectile's motion in a grazing collision. 
Thus, the collisional outcomes of differentiated icy planetesimals are similar to those of pure icy planetesimals. 
In fact, a hit-and-run occurs for the same impact velocity regardless of the existence of a core.
Note that since pure icy planetesimals and pure rocky planetesimals have similar critical impact velocities, even if the core--mantle ratio changes, the critical impact velocity may not change significantly.

\subsubsection{Comparison of All Merging Criteria}

\begin{figure}  
\includegraphics[width=\textwidth]{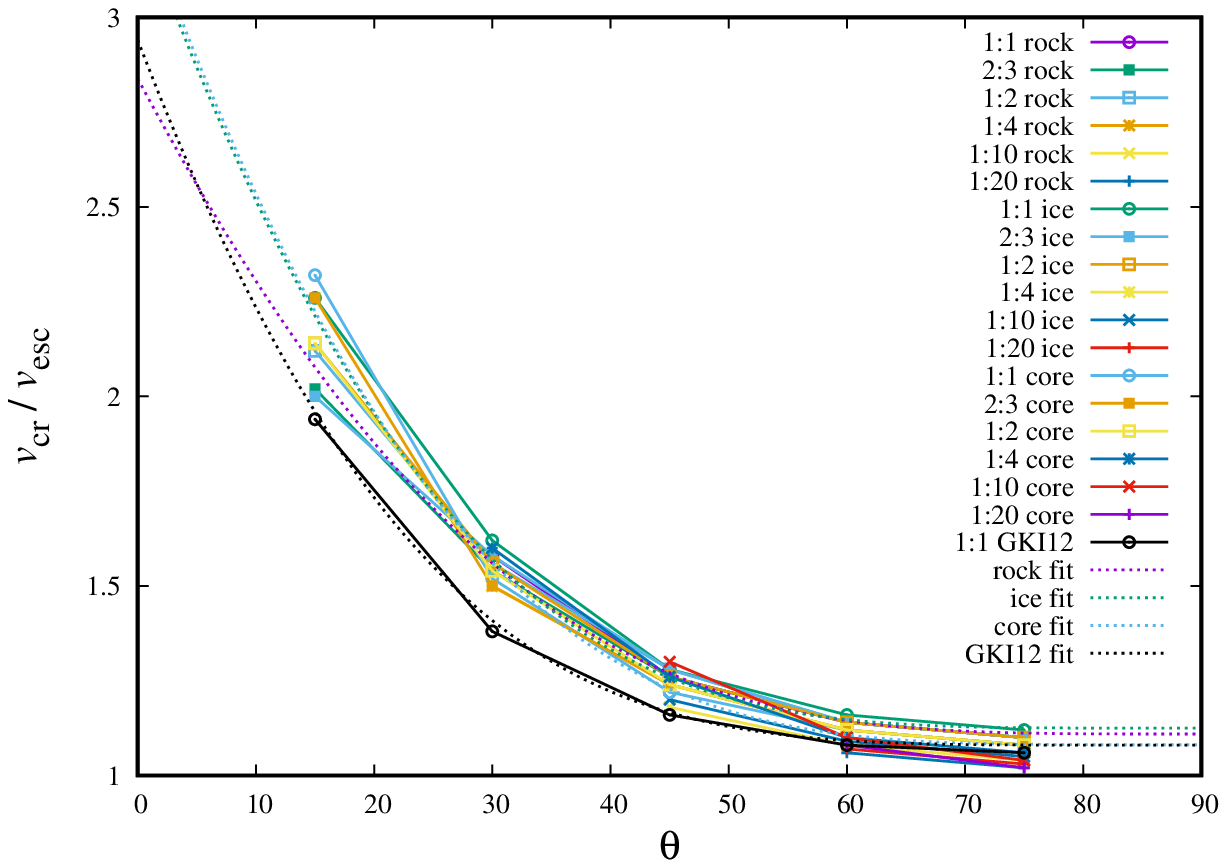}
\caption{
As Figure \ref{fit}, comparing all types of planetesimals. For each symbol, the color indicates the composition (pure rocky, pure icy, and differentiated icy planetesimal).
The dotted curves represent the fitting formula Eq. (\ref{fitting}) of all compositions for $\gamma = 1$.
The different colors correspond to the different compositions.
The black circles indicate our results calculated with the same conditions as in GKI12.
The dotted black curve represents the fitting formula Eq. (\ref{fitting}) of GKI12 for $\gamma = 1$.
\label{compare}}
\end{figure}

The critical impact velocity of all planetesimal types calculated in this study is plotted against the impact angle for $\gamma = 1$ together with the fitting formula of GKI12 where $c_1 = 2.43$, $c_2 = -0.0408$, $c_3 = 1.86$, $c_4 = 1.08$, and $c_5 = 2.5$, in Fig. \ref{compare}. 
We find that the merging criteria of planetesimals are almost independent of their composition and internal structure.

Although the difference in the critical impact velocity is small, the composition has more of an effect than the internal structure.
The maximum difference in the critical impact velocity between the different compositions is $0.2v_{\rm esc}$ at $\theta = 15^\circ$, while that between the differentiated and undifferentiated planetesimals is $0.06v_{\rm esc}$ at $\theta = 75^\circ$. 
The composition determines the pressure at a collision and thus the critical impact velocity.
The existence of a core does not affect the pressure at a collision since the mantle is responsible for the pressure. 

GKI12 presented the merging criteria of rocky protoplanets with an iron core and a rocky mantle. 
Our results for the same initial conditions are shown in Figure \ref{compare} and are consistent with the fitting formula of GKI12.
The critical impact velocity has a maximum difference from that of icy differentiated planetesimals of about $0.2v_{\rm esc}$ around $30^\circ$.
This difference in the critical impact velocity is consistent with that between pure icy and rocky planetesimals, which shows again that it is the mantle composition that determines the critical impact velocity. 

For the realistic impact parameters of planetesimal accretion $v_{\rm imp} \simeq 1.0v_{\rm esc}$ and $\theta \simeq 45^\circ$, 
the difference in the critical impact velocity between the rocky planetesimals and the differentiated rocky protoplanets is $\sim 0.15v_{\rm esc}$ (Fig. \ref{compare}). 
This difference affects the accretion efficiency of planetesimals. 

Since a differentiated rocky protoplanet with an iron core in GKI12 is not largely deformed by the tidal force due to the existence of the core, the collision occurs at the same collision angle as for our study. 
In this case $v_{\rm cr}$ increases as $\gamma$ decreases.
This tendency is also clearly shown in our results for the differentiated icy planetesimals for $\theta = 45^\circ$ in Figure \ref{differentiated}.
Thus, the results for the differentiated icy planetesimals are consistent with GKI12. 
However, due to tidal deformation, the dependence of $v_{\rm cr}$ on $\gamma$ for undifferentiated icy planetesimals is different from that of differentiated icy planetesimals.
The effect of tidal deformation is stronger for undifferentiated bodies and becomes prominent as $\gamma$ increases.
This increases the collisional cross-section of undifferentiated icy planetesimals and $v_{\rm cr}$ becomes higher than that of differentiated icy planetesimals.
The dependence of $v_{\rm cr}$ on $\gamma$ of the undifferentiated rocky planetesimals is weaker than that of the rocky protoplanets in GKI12 due to the tidal deformation.

It should be noted that if we include the cohesion and/or friction in the planetesimal model, they prevent tidal deformation and
planetesimals become harder than our fluid-model. 
The colliding planetesimals may not deform significantly when they approach.
In that case, $v_{\rm cr}$ depends on $\gamma$ significantly and $v_{\rm cr}$ increases with decreasing $\gamma$.

The total mass independence of $v_\mathrm{cr}$ allows us to compare our results to \cite{2019ApJ...875...40C} and \cite{2020ApJ...892...40G}, which studied the collisional outcomes of planetary embryos using machine learning with SPH collision experiment results. 
Their model is basically the same as ours and their results generally agree with our results. 
The boundary between hit-and-run and merging of colliding objects in \cite{2019ApJ...875...40C} is consistent with our results.
In fact, the boundary between "Graze and Merge" and "HnR" in their Fig.4 almost perfectly matches the boundary in Fig.10.

We obtained the merging criteria of collisions for undifferentiated icy, undifferentiated rocky, and differentiated icy planetesimals. 
They have small but not negligible differences.
The merging criteria help us understand the realistic accretion process of planetesimals over a whole protoplanetary disk. 

\begin{table}
    \centering
\begin{tabular}{ |p{5.0cm}||p{1.6cm}|p{1.6cm}|p{1.6cm}|p{1.6cm}|p{1.6cm}|  }
 \hline
 \multicolumn{6}{|c|}{Coefficients of Equation (1)} \\
 \hline
 Planetesimal Types & $c_1$ & $c_2$ & $c_3$ & $c_4$ & $c_5$\\
 \hline
 Undifferentiated Icy   &  -0.232 &  -0.114 &  2.14 &  1.12 &  2.27\\
 Undifferentiated Rocky & -0.00863 & -0.107 & 1.73 & 1.11 & 1.94\\
 Differentiated Icy     & 0.313 & -0.0249 &  2.21    & 1.08 & 2.21\\
 Differentiated Rocky (GKI12)  & 2.43 & -0.0408 &  1.86    & 1.08 & 2.50\\
 \hline
\end{tabular}
    \caption{Fitting coefficients of equation (1).}
    \label{table1}
\end{table}

\section{Summary and Discussion}

We performed SPH collision experiments of rocky/icy planetesimals and obtained the critical impact velocity that distinguishes merging from hit-and-run outcomes.
We found that the critical impact velocity does not depend on the total mass of the colliding planetesimals in the gravitational regime. 
On the other hand, the critical impact velocity strongly depends on the impact angle and weakly depends on the mass ratio (projectile mass divided by target mass) of the planetesimals. 
The critical impact velocity decreases as the impact angle of the colliding planetesimals increases.
The critical impact velocity normalized by the escape velocity of the planetesimals decreases as the mass ratio decreases, particularly so in grazing collisions.
We found that planetesimal deformation changes the collision angle, and less energy dissipation occurs for smaller mass ratios. Thus, collisions with smaller mass ratio result in hit-and-run outcomes.
The planetesimal composition has a stronger affect the merging criteria than the existence of a core.
The composition determines the pressure at the collision and thus the critical impact velocity. The existence of a core does not affect the critical impact velocity since the mantle determines the pressure. 
However, the dependence of the merging criteria on the composition and the internal structure is relatively weak.
We derived a fitting formula for the merging criteria of undifferentiated rocky/icy planetesimals and differentiated icy planetesimals (Eq. \ref{fitting}).

These results are obtained under the assumption of fluid-like
planetesimals without material strength, internal friction, porosity
and internal structure. The material strength is not negligible for
small planetesimals in particular for monolithic (non-porous) ones.
The collisional outcomes depend on the material strength of monolithic
planetesimals \citep{2019Icar..317..215J}. Planetesimal deformation is hindered
by the material strength, which makes it harder for the impact energy
to dissipate, leading to bouncing back. If the internal friction is
considered, planetesimals become harder and defrom less than our model (Jutzi 2015).
In this case the critical impact velocity can be higher because
the energy is also dissipated by the friction.
For the porous and rubble pile planetesimals the collisional outcomes
depend on the porosity and internal structure.
Generally they have higher energy dissipation and deformability,
making them easy to merge. These effects must be considered in a
future work of merging criteria of planetesimals.

Fragmentation is important for high-speed collisions.
We see that the mass of fragments increases with the impact velocity, which is consistent with previous studies \citep[e.g.,][]{2012ApJ...745...79L, 2017Icar..294..234G}. 
However, the mass of fragments in our study is larger than that of the higher resolution simulations.
For example, in a collision of planetesimals of radius 100 km (mass  $7.0\times 10^{-7}M_{\oplus}$) with the impact parameters $\theta = 30^\circ$ and $v_\mathrm{imp} = 3.0v_{\rm esc}$, the total mass of the fragments is about 0.01 times of the total mass of the system \citep{2017Icar..294..234G}, while in our simulation it is about 0.3 times the total mass of the system.
The difference is mainly due to the number of SPH particles. 
\citet{2017Icar..294..234G} used $10^7$ particles for a target, while we used just 10,000 for the smallest target. 
Although 10,000 SPH particles is sufficient for investigating the merging criteria, this does not allow the fragmentation to be correctly followed \citep{2015Icar..262...58G}.
For an extremely small mass ratio, projectiles are broken into fragments by tidal deformation and collision and some of the fragments fly away. 
Evaluating the mass accretion of a projectile on a target for an extremely small mass ratio is difficult since the resolution of the calculation is not high enough to simulate realistic fragmentation.
Thus, our merging criteria might be not realistic for extremely small mass ratios.
We need to consider the uncertainty when investigating the planetesimal accretion process.
To understand the collisional outcomes with realistic fragmentation, future work will require extremely high-resolution SPH collision experiments.

After protoplanets are formed, the protoplanets perturb the remaining planetesimals. 
In the oligarchic growth stage, the impact velocity of planetesimals is on the order of their escape velocity.
According to the merging criteria obtained in this study, it is possible that a few tens of percent of colliding planetesimals will bounce back, which affects the accretion time of the planetesimals.
In other words, hit-and-run collisions increase the accretion time and may change the mass distribution of planetesimals from that of the perfect merging case.
Thus, the merging criteria of planetesimal collisions must be considered when investigating the accretion process of planetesimals.
In a future paper, we plan to incorporate the merging criteria of planetesimals into $N$-body simulations and study a more realistic planetesimal accretion process taking into account merging and hit-and-run outcomes.

\acknowledgements

The numerical computations used in this study were carried out on ATERUI II (Cray XC50) at the Center for Computational Astrophysics, National Astronomical Observatory of Japan. 
T. S. wishes to thank Hidenori Genda, Akihiko Fujii, and Benjamin Wu for their helpful advice.
This work was supported by JSPS KAKENHI Grant Number 18H05438.

\appendix

\section{Tillotson Equation of State}

The Tillotson EOS takes three different formulas for pressure calculation depending on the density, $\rho$, and the specific internal energy, $u$.
Hereafter, $\rho_0$, $u_0$, $a$, $b$, $A$, $B$, $u_{\rm cv}$, $u_{\rm iv}$, $\alpha$, and $\beta$ are material parameters.
We use the parameter sets of basalt and ice for planetesimals listed on Table II of page 7 in \cite{1999Icar..142....5B}.
In the condensed ($\rho > \rho_0$) or cold state ($u < u_{\rm iv}$) region the Tillotson EOS is given as
\begin{equation}
    p_{\rm co} = \left(a + \frac{b}{\frac{u}{u_0 \eta^2} + 1 }\right)\rho u + A\mu + B\mu^2 ,
    \label{condensed}
\end{equation}
where $\eta = \rho / \rho_0$ and $\mu = \eta - 1$.
In the expanded hot state ($\rho < \rho_0$ and $u > u_{\rm cv}$) the Tillotson EOS takes the form:
\begin{equation}
    p_{\rm ex} = a\rho u + \left\{ \frac{b\rho u}{\frac{u}{u_0 \eta^2} + 1} + A \mu {\rm exp}\left[-\beta\left(\frac{1}{\eta}-1\right)\right]\right\} \times {\rm exp} \left[-\alpha \left(\frac{1}{\eta}-1\right)^2\right].
\end{equation}
In the intermediate region ($u_{\rm iv} < u < u_{\rm cv}$ and $\rho < \rho_0$), a smooth transition between the above two states occurs.
Thus, following \citet{1986Icar...66..515B}, we interpolate the pressure by using $p_{\rm co}$ and $p_{\rm ex}$:
\begin{equation}
    p_{\rm tr} = \frac{(u - u_{\rm iv})p_{\rm ex} +(u_{\rm cv} - u)p_{\rm co}}{u_{\rm cv} - u_{\rm iv}}.
\end{equation}

\end{document}